\documentclass[twocolumn,showpacs,floatfix,prl]{revtex4}
\usepackage{amssymb}
\usepackage{graphicx}
\usepackage{dcolumn}
\usepackage{bm}
\usepackage[usenames]{color}

\begin{document}

\title{Interplay between charge-lattice interaction and strong electron correlations
in cuprates: phonon anomaly and spectral kinks}
\author{G.~De~Filippis$^1$, V.~Cataudella$^1$, R. Citro$^2$, C.A. Perroni$^1$, A.~S.~Mishchenko$^{3,4}$ and
N.~Nagaosa$^{3,5}$}
\affiliation{$^1$SPIN-CNR and Dip. di Scienze Fisiche - Universit\`{a} di Napoli
Federico II - I-80126 Napoli, Italy \\
$^2$ CNISM and Dip. di Fisica ``E. R. Caianiello''- Universit\`{a} di Salerno, I-84100 Salerno, Italy
\\
$^3$ Cross-Correlated Materials Research Group (CMRG), ASI, RIKEN,
Wako 351-0198, Japan
\\
$^4$RRC ``Kurchatov Institute'' - 123182 - Moscow - Russia\\
$^5$Department of Applied Physics, The University of Tokyo,
7-3-1 Hongo, Bunkyo-ku, Tokyo 113, Japan}


\begin{abstract}

We investigate the interplay between strong electron correlations and
charge-lattice interaction in cuprates.
The coupling between half breathing bond stretching
phonons and doped holes in the $t$-$t'$-$J$ model is studied by limited phonon
basis exact diagonalization method.
Nonadiabatic electron-phonon interaction leads to
the splitting of the phonon spectral function at half-way
to the zone boundary at $
\vec{q}_s=\{(\pm \pi/2,0), (0, \pm \pi / 2) \}$ and to low energy kink feature in the electron dispersion,
in agreement with experimental observations.
Another kink due to strong electron correlation effects is observed at higher energy, depending on 
the strength of the charge-lattice coupling.

\end{abstract}

\maketitle

There is a growing
confidence that strong electron-phonon interaction (EPI) manifests itself
both in vibrational \cite{PintsRev} and electronic
\cite{GunRev} spectra of cuprates.
The most puzzling feature of the cuprate phonon spectra
is the anomaly of the half breathing bond stretching (HBBS)
phonon occurring at half-way to
the Brillouin zone (BZ) boundary in the [100]-direction, while the
most debating feature of the electronic spectra are the
kinks observed in Angle Resolved Photoemission Spectra (ARPES).

It was recently realized in the experimental community that
it is highly important to measure both phonons spectra and ARPES
on the same sample just to verify possible link between
HBBS phonons and ARPES \cite{Graf08}.
The results of the above studies support close connection
between HBBS phonon anomaly and lowest energy kink in ARPES \cite{Graf08}
and, thus, a model describing both
anomalies within the same approach is strongly needed.

In the present Letter we study the low density
limit (one hole on $4 \times 4$ lattice)
of the extended $t$-$t'$-$J$ model where
holes are coupled to HBBS phonons \cite{bulut}.
In order to calculate the phonon spectral function (PSF)
and hole spectral function (HSF), we generalize
a recently introduced approach \cite{OurCM}, based on limited phonon
basis exact diagonalization (LPBED),
without adopting self-consistent Born (SCBA)
and spin wave approximations \cite{Kane,Liu_92}.
This method treats the non-adiabatic effects of the quantum
phonon very effectively without approximations
for the magnetic degrees of freedom.
Because of exponential growth of the basis
with size of the system, the $4 \times 4$ lattice
has considerably denser quantum states
than the smaller $\sqrt{10}\times\sqrt{10}$ system \cite{Fehske},
so that it is possible to resolve fine structure
of the PSF and HSF.
For the first time we are able to reveal the
shape of the PSF while previous
studies \cite{KhaPho,GuPho,KhaBS}
were restricted to at most the second
moment of the response.

We show that EPI can lead to the splitting
of the PSF at half-way
to the BZ boundary in the $[100]$ direction.
We demonstrate that the splitting can be
easily smeared out by a very small broadening
and, thus,
the mysterious evasive behavior of double-peak
structure can be attributed to tiny
variations of chemical composition, crystal
quality and/or experimental setup.
We argue that the splitting is a rather general
phenomenon arising when the phonon branch interacts
with a rather soft electronic excitation and
we show that the HBBS phonon anomaly is linked
to the lowest energy kink observed in the ARPES.
Finally we emphasize that the same model supports
the spectral kink at higher energy referred to
colloquially as the waterfall \cite {Graf1}.

McQueeney et al.~\cite{McQu99}
observed anomalous lineshape of HBBS phonons
around ${\bf q}=(\pi/2,0,0)$ and reported its strong temperature dependence.
Subsequent studies confirmed that
the HBBS phonon anomaly is due to the coupling of the
HBBS phonons to the doped carriers. Indeed,
the anomaly is absent in undoped compounds
\cite{PintsRev} and the hardening of the phonon spectra
with heating \cite{McQu99,Rez07} excludes such
sources of anomaly as anharmonicity or structural
inhomogeneity.
It is also agreed that HBBS phonon anomaly
is located around ${\bf q}=(\pi/2,0,0)$ at any doping.
On the other hand, there is a
controversy on the relation between the HBBS phonon anomaly
and the lowest energy kink in the ARPES.
This relation is often denied although
the measurements of the phonon spectra
and ARPES on the same sample of
Bi$_2$Sr$_{1.6}$La$_{0.4}$Cu$_2$O$_{6+x}$
suggest that the softening of the HBBS phonon
mode matches the energy and momentum
of this kink \cite{Graf08}.
Another much debated question is
the interpretation of the
PSF structure at HBBS phonon anomaly.
Actually the phonon peaks at $\vec{q}_s$
are poorly defined and two scenarios
are possible. According to the single-branch interpretation
(SBI) there is softening and a very large intrinsic linewidth
of a single peak near ${\bf q}=(\pi/2,0,0)$.
On the contrary, the double-branch interpretation
(DBI) implies the presence of two peaks:
the first one is located at the
unrenormalized energy and the second one is
broad and soft.
The pioneering study in Ref.~\cite{McQu99} was followed by
\cite{Pints99} whose results supported SBI.
Next studies of similar material
La$_{1.875}$Ba$_{0.125}$CuO$_4$
led authors to conclusion about relevance of
DBI \cite{Rez06}. Likewise, measurement of HBBS
in YBa$_2$Cu$_3$O$_{6+x}$ was interpreted in the framework of
DBI \cite{Strec08}. Recently a "contamination hypothesis" \cite{RezPrivate}
has been suggested, claiming
that the SBI is correct because the inelastic neutron
scattering measurement picks up the
intensity from the k-vicinity $(\pi/2,k,0)$
of ${\bf q}=(\pi/2,0,0)$ point and, thus, the "normal"
component comes from transverse phonons.
However, the above reasoning contradicts the results
obtained on La$_{2-x}$Ba$_{x}$CuO$_{4+\delta}$
at $x=0.14 \pm 0.01$ \cite{dAstuto08} by
inelastic x-ray scattering technique which
has higher momentum resolution than that in
neutron scattering measurements.
"Contamination hypothesis" implies that the "normal"
component must disappear by improving momentum
resolution.
Instead,
the results of Ref.~\cite{dAstuto08} support DBI.
One can guess that the SBI vs DBI depends on the
chemical composition \cite{RezPrivate},
Ba-doped \cite{Rez06,dAstuto08} vs Sr-doped
\cite{Pints99,Rez07} compound.
Besides, since the softening
is very sensitive to the broadening
caused by decay channels and/or experimental resolution,
SBI vs DBI must be strongly dependent on the compound,
sample quality, and experimental setup.

We will show that many of the questions discussed above can be understood
within the $t$-$t'$-$J$ model including the coupling with HBBS.
On the contrary density functional theory
calculations \cite{Giu08,Bohn03} do not
predict any HBBS phonon anomaly \cite{Rez08}.
Other approaches, especially
those associating the anomaly with stripes, have
difficulty with the position of
HBBS phonon anomaly.
Scenario suggested in Ref.~\cite{KohnAno} relates
the HBBS softening to the Kohn anomaly at double Fermi
momentum $2k_F$ along the Fermi surface of stripes
\cite{KohnAno}.
Here, in contrast to experiment, the softening must
be $\theta$-independent.
In a different theoretical proposal, HBBS
anomaly is associated with stripe mediated collective
charge excitations \cite{Kaneshita02} or incommensurate
low energy spin-fluctuations \cite{Kivel03}.
Within these scenarios, in contrast with experiment
\cite{Rez06,dAstuto08,Graf07,Rez07,Uchi04},
wave vector of phonon anomaly strongly depends on
doping level. Finally we note that phonon softening
\cite{KhaPho,GuPho}, broadening \cite{KhaPho},
and correct position of the anomaly \cite{KhaBS}
have already been qualitatively explained by the coupling
between phonons and density response of the $t$-$J$ model
although none of those studies considered the shape of the
phonon spectral function.

The Hamiltonian of the $t$-$t'$-$J$-Holstein model in 2D is the sum of the electronic part and
hole-phonon coupling Hamiltonian
\begin{eqnarray}
H_{tt^{'}J}=&& -t \sum_{i,\delta,\sigma} c_{i+\delta,\sigma}^{\dagger}c_{i,\sigma}
-t' \sum_{i,\delta^{'},\sigma} c_{i+\delta^{'},\sigma}^{\dagger}c_{i,\sigma} \nonumber \\
&& + \frac {J}{2} \sum_{i,\delta} S_{i+\delta} S_{i}
- \frac {J}{8} \sum_{i,\delta} n_{i+\delta} n_{i} \; ,
\end{eqnarray}
\begin{eqnarray}
H'= \omega_0 \sum_{q,\mu}
a^{\dagger}_{q,\mu} a_{q,\mu}+
\sum_{i,q,\mu} \left( M_{q,\mu} e^{i \vec{q}\cdot \vec{R}_i} (1-n_i)
a_{q,\mu} + H.c.\right).
\nonumber
\end{eqnarray}
Here $J$ is the exchange interaction constant of the
spin-spin interaction,
$t$ and $t'$ are hopping amplitudes to nearest and next
nearest neighbors.
At site i (Cu atoms), $S_i$ is the $\frac {1} {2}$-spin operator,
$c_{i, \sigma}$ is the fermionic operators in the
space without double occupancy, and
$n_i$ is the number operator.
$a_{q,\mu}$ is the phonon annihilation operator
with momentum $q$ and $\mu=x$ or $y$ indicates the longitudinal
polarization of the oxygen vibrations along the direction of the
nearest neighbor Cu atoms (Oxygen atoms are located at $\vec{R}_i+a/2\mu$,
$a$ being the lattice parameter).
$M_{q,\mu}$ is the matrix element of the EPI:
$M_{q,\mu}=g\omega_0/\sqrt{N} 2i \sin(q_{\mu}/2)$ where $N$
is the number of lattice sites and $\omega_0$ is the frequency of dispersionless
optical phonon.
The strength of EPI is characterized by dimensionless coupling
constant $\lambda=\sum_{\vec{q},\mu} |M_{q,\mu}|^2/4 \omega_0 t$.
We chose parameters which correspond to hole doped cuprates:
$J=0.4t$, $t'=-0.25t$, $\omega_0=0.15t$, and set $t=a=\hbar =1$.
The PSF is expressed as
\begin{eqnarray}
D(\vec{q},\omega+i\eta) = - \frac {1} {\pi} \Im \frac {1} {D^{-1}_{0}(\vec{q},\omega+i\eta)-\Sigma(\vec{q},\omega+i\eta)},
\end{eqnarray}
where the self-energy $\Sigma(\vec{q},\omega+i\eta)$ is given by
\begin{eqnarray}
\Sigma(\vec{q},\omega+i\eta)=
\frac  {|M_{q,x}|^2 \Pi(\vec{q},\omega)}{(1+|M_{q,x}|^2 \Pi(\vec{q},\omega)/D^{-1}_{0}(\vec{q},\omega+i\eta))}.
\end{eqnarray}
Here $D_{0}(\vec{q},\omega+i\eta)$ is the bare phonon Green function and $\Pi(\vec{q},\omega)=
P(\vec{q},\omega+i\eta)+P(\vec{q},-\omega-i\eta)$ is the polarization insertion with
\begin{eqnarray}
P(\vec{q},\omega+i\eta) = \left\langle \psi_0 \right | O^{\dagger} \frac {1} {\omega+i\eta-H+E_0}
O \left | \psi_0 \right\rangle.
\label{pi}
\end{eqnarray}
We choose the ground state (GS) $\left | \psi_0 \right\rangle$ as
a linear superposition with equal weights of the 4 degenerate states corresponding to
$\vec{k}=(\pm \frac {\pi} {2}, \pm \frac {\pi} {2})$ with
energy $E_0$, $\eta$
is a broadening factor that shifts the poles of $D(\vec{q},\omega)$ in the complex plane, and
$O = \sum_{i} e^{i \vec{q}\cdot \vec{R}_i} (1-n_i)$.
The increase of broadening factor $\eta$ has the
physical meaning
of phonon damping or limited experimental resolution.

The ground state $\left | \psi_{GS} \right\rangle$
and the function $P(\vec{q},\omega+i\eta)$ are
obtained by modified \cite{dagotto} and standard
Lanczos methods, respectively, within the LPBED method \cite{OurCM} introduced for the
$t$-$t'$-$J$-Holstein model.

\begin{figure}
\flushleft
        \includegraphics[scale=0.8]{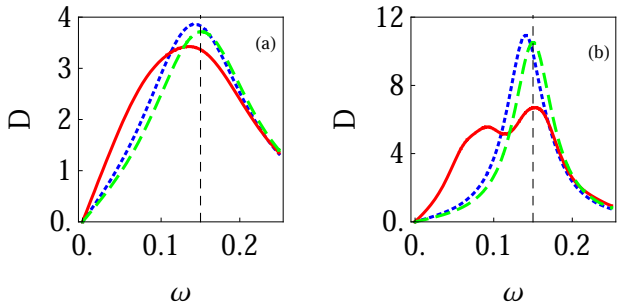}
        \caption{(Color online) (a) The PSF for $\lambda=0.3$ at
three different values of $\vec{q}$ ($\vec{q}=(0,0)$ (dashed green line), $\vec{q}=(\pi/2,0)$ (solid red line),
and $\vec{q}=(\pi,0)$ (dotted blue line)) with $\eta=0.08$; (b) the PSF
with a smaller broadening factor $\eta$, $\eta=0.03$. The vertical lines indicate the bare phonon frequency $\omega_0$.}
\label{fig1}
\end{figure}

\begin{figure}
        \includegraphics[width=85mm,height=40mm]{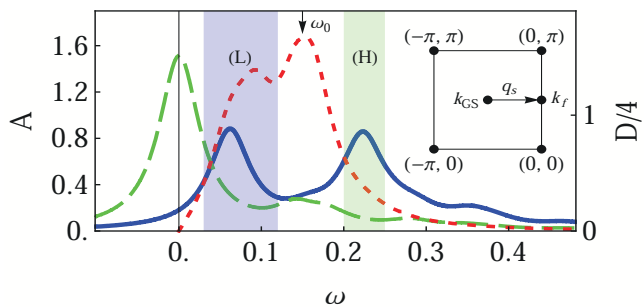}
        \caption{(Color online) HSF $A(\vec{k},\omega)$,
for $\lambda=0.3$ at two different values of $\vec{k}$ ($\vec{k}=(\pi/2,\pi/2)$ (dashed green line),
$\vec{k}=(0,\pi/2)$ (solid blue line)) with $\eta=0.03$. The energy is measured with respect to the GS energy.
The phonon Green function at $\vec{q}=(\pi/2,0)$ (with the same value of $\eta$)
is plotted too for comparison (red dotted line).
In the inset sketch of one of transitions involved in the PSF calculation.}
\label{fig2}
\end{figure}

In Fig. 1 we plot the PSF
$D(\vec{q},\omega)$ for different wavevectors along the $(1,0)$
direction in the BZ.
For the chosen values of the model parameters the system
undergoes, in agreement with \cite{AlltJ}, a crossover
towards strong EPI regime for
$\lambda_c \simeq 0.5$. The anomalous softening of the phonon mode
at $\vec{q}_s=(\pi/2,0)$ is already observed at moderate
values $\lambda < \lambda_c$
of the hole-phonon coupling ($\lambda=0.3$).
We stress that for a large value of broadening factor $\eta=0.08$
the phonon peak softens and broadens at $\vec{q}_s$ (Fig. 1a),
supporting, thus, SBI. On the other hand,
reducing the broadening factor to $\eta=0.03$ two-peak structure
turns out(Fig. 1b), that is in agreement with the
experimental observations reported in Ref.\cite{Rez06,dAstuto08}.

To clarify the physical nature of the splitting, we study the
hole spectral weight function $A(\vec{k},\omega)$
at the wavevectors $\vec{k}_f$ that are reached starting from
GS wavenumber at $(\pm \pi/2, \pm \pi/2)$ through
phonon momentum $\vec{q}_s=(\pi/2,0)$.
For example, as shown in Fig.2,
by starting from $\vec{k}=(-\pi/2,\pi/2)$ we get
$\vec{k}_f=(0,\pi/2)$
after absorbing (emitting) a phonon with momentum $-\vec{q}_s$ ($\vec{q}_s$).
The lowest energy peak in $D(\vec{q}_s,\omega)$
(dotted red curve in Fig. 2) is close to
the lowest energy peak of $A(\vec{k}_f,\omega)$
(solid blue curve in Fig. 2).
Measuring energy from the GS, denoted by
vertical line in Fig.~2, one can see that
the lowest peaks in phonon and hole spectral
functions are softer than the phonon energy $\omega_0$=0.15t.
We note that the significant renormalization of the hole
spectral weight, with respect to GS one,
indicates strong
coupling between phonon at $\vec{q}_s$ and
hole at $\vec{k}_f$.
The high energy peak in $D(\vec{q}_s,\omega)$
is located at energy close to $\omega_0$, and no peak
at the same energy is found in $A(\vec{k},\omega)$.
It is shown below that it is due to
the phononic nature of the high energy
resonance of the PSF.

To give a simple explanation of the above scenario one has to
realize that the excited electronic state
$\left | \psi \right\rangle ^0_{(0,\pi/2)} \left | 0 \right\rangle$
with the momentum $(0,\pi/2)$ and without phonons
is linked by the matrix elements of
EPI with a group of 8 degenerate states
$\left | \psi \right\rangle ^0_{(\pm \pi/2,\pm \pi/2)}
a^{\dagger}_{q,\mu}\left | 0 \right\rangle$
where the electronic subsystem is in the GS and
one phonon is excited.
The momentum conservation
$(\pm \pi/2,\pm \pi/2)+\vec{q}=(0,\pi/2)$
determines the phonon momenta $\vec{q}$.
The energies of the electronic subsystem at corresponding
momenta are $\epsilon^0(0,\pi/2)$ and $\epsilon^0(\pi/2,\pi/2)$,
respectively.
Then, choosing a set of the model
parameters appropriate for cuprates, the resonance relation
$\epsilon^0(0,\pi/2) \approx \epsilon^0(\pi/2,\pi/2) + \omega_0$
is satisfied.
Hence, even small matrix elements induce strong
effects both in electronic and vibrational subsystems.
Analytic diagonalization of this degenerate $9 \times 9$
matrix gives 9 levels.
The lowest state $L$ has energy below $\omega_0$ and has
components on both one-phonon and
zero-phonon states.
This state corresponds to the peaks of electronic
$A(\vec{k}=(0,\pi/2),\omega)$ and bosonic
$D(\vec{q}=(\pi/2,0),\omega)$
spectral functions at $\omega<0.1$
(see the area around letter $L$ in Fig.~2).
In the highest energy range, (see the area around letter
$H$ in in Fig.~2), both spectral functions
collect contributions at $\omega \ge 0.2$.
However, the $H$-state is of predominantly
electronic origin with a large peak in
the hole spectral function while it generates
a weak structure in the PSF,
hardly observable in the numerical data.
The last 7 degenerate levels with energy
$\epsilon^0(\pi/2,\pi/2)+\omega_0$ have no projection on the
vacuum boson state: this explains why the hole spectral weight function does not show any peak at this energy,
at a distance $\omega_0$ from GS energy.
The energies of all levels of the above simple analytic
solution are in qualitative agreement with that provided
by the numeric LPBED method in Fig.~2.
So, we conclude that the doubling of the phonon peak is due to coupling between holes and lattice,
that lifts a degeneracy and produces one additional state with energy less than $\omega_0$.

Generically, the EPI involving electronic states
with energies considerably larger than the phonon
frequency does not lead to any splitting since
the adiabatic approximation is valid and electron
produces only a renormalization of the
adiabatic potential \cite{BornOpp27}.
On the other hand, when the electronic excitation
is soft and its energy is comparable with phonon
frequency, the nonadiabatic corrections play
a crucial role \cite{KiMi93} and one can observe
exotic spectral functions as it is seen in Fig.~1b and
in experiments on cuprates.

\begin{figure}
\flushleft
        \includegraphics[scale=0.8]{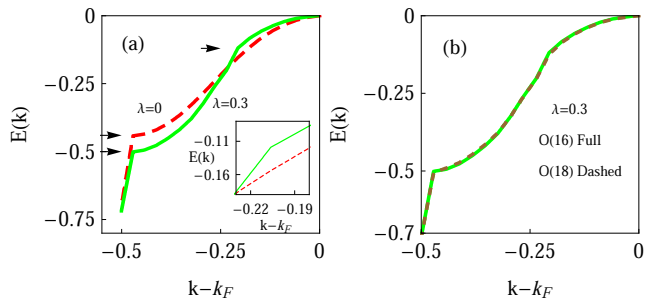}
        \caption{(Color online) (a) Energy vs $(\vec{k}-\vec{k}_F)/\pi$
dispersion relations along nodal direction from
$\vec{k}_F$ to $\Gamma$ point at $\lambda=0.3$ (solid) and $\lambda=0$ (dashed).
Arrows mark kink positions. (b) Dispersion relation as a function of oxygen isotope exchange ($^{16}$O (solid) and $^{18}$O (dashed)) at $\lambda=0.3$.}
\label{fig3}
\end{figure}

The clear connection, established by the above
considerations, between phonon at $\vec{q}_s$ and
hole at $\vec{k}_f$ encouraged us to
search for an interplay
between the HBBS phonon anomaly and the lowest energy kink in the
dispersion of quasiparticles in cuprates
\cite{Lanzara}. In Fig.3a we plot hole dispersion derived from
momentum distribution curves, i.e. we find the $\vec{k}$ for which
the hole Green function $A(\vec{k},\omega)$
has maximum at fixed $\omega$ \cite{end}. The energy, measured
from GS, is plotted versus the wavenumber $\left( \vec{k}-\vec{k}_F \right )$,
where $\vec{k}_F$ is the momentum corresponding
to the minimum of the hole dispersion relation.
We find that, when $E_k-E_{k_{F}}$ is about
$\omega_0=0.15$, the curve exhibits a slope change related to
the coupling between the bare phonon energy and the
electronic band in the bare $t$-$t'$-$J$ model. This kink does not appear
in absence of the hole-phonon coupling. On the other hand, we obtain
another kink at higher energy (see Fig.3a) \cite {Graf1}. This distinctive feature is related to
the strong electron correlations since it is observed also
in the bare $t$-$t'$-$J$ model. At $\lambda=0$ the kink is located around the exchange interaction
energy $J$, in agreement with results by Chakraborty et al. \cite{phillips} within the Hubbard model.
Furthermore, the kink is shifted at higher energy by increasing the hole-phonon coupling. We also
investigated the effect of the oxygen isotope
substitution \cite{OurCM} on ARPES.
The results (Fig.3b) show a negligible (below $0.01t$)
shift upon oxygen isotope exchange,
that is in agreement with recent experimental observations by Douglas et al. \cite{Dessau}.
All these data point out that the scenario based on the interplay
between strong electron correlations and hole-phonon interaction
is able to capture many physical distinctive features
of high temperature superconductors.

In conclusion, we showed
that the EPI,
in the presence of strong correlations,
can lead to the splitting
of the phonon spectral function at half-way
to the BZ boundary in the $[100]$
direction. We demonstrated that the splitting can be
easily smeared out by very small broadening
of the eigenstates. The same
physical mechanism can explain both
HBBS phonon anomaly and lowest energy kink in the ARPES.
Finally we found that the isotope effect on ARPES is negligible in accordance
with experiment. These results support the claim that strong electron correlations
and charge lattice interaction are crucial in understanding
cuprate experimental features.

Work supported by RFBR 07-02-0067a (A.S.M.), and Grant-
in-Aids No. 15104006, No. 16076205, No. 17105002, No.
19048015, and NAREGI Japan (N.N.); G.D.F., V.C. and C.A.P.
received financial support from MIUR-PRIN 2007 under Prot.
No. 2207FW3MJX003.

\end{document}